\documentclass[11pt]{article}
\usepackage[utf8]{inputenc}

\usepackage{threeparttable}
\usepackage{amsxtra}
\usepackage{url}
\usepackage{amsthm}
\usepackage{amssymb}
\usepackage{amsfonts}
\usepackage{graphicx}    
\usepackage{rotating}
\usepackage{color}
\usepackage{epsfig}
\usepackage{subfigure}   
\usepackage{verbatim}
\usepackage[numbers]{natbib}
\usepackage{algorithm,algcompatible}
\algnewcommand\INPUT{\item[\textbf{Input:}]}
\algnewcommand\OUTPUT{\item[\textbf{Output:}]}%
\usepackage{float}
\usepackage[printonlyused]{acronym} 
\usepackage{setspace}  
\usepackage{mathtools}
\usepackage{geometry}     
\usepackage{caption}
\usepackage{natbib}

\usepackage{indentfirst} 
\usepackage{color}
\usepackage[dvipsnames]{xcolor}
\usepackage{tikz}

\DeclareRobustCommand{\blackline}{\raisebox{2pt}{\tikz{\draw[-,black!40!Black,solid,line width = 0.9pt](0,0) -- (5mm,0);}}}
\DeclareRobustCommand{\blueline}{\raisebox{2pt}{\tikz{\draw[RoyalBlue,solid,line width = 0.9pt](0,0) -- (5mm,0);}}}
\DeclareRobustCommand{\orangeline}{\raisebox{2pt}{\tikz{\draw[BurntOrange,solid,line width = 0.9pt](0,0) -- (5mm,0);}}}
\DeclareRobustCommand{\greenline}{\raisebox{2pt}{\tikz{\draw[Green,solid,line width = 0.9pt](0,0) -- (5mm,0);}}}
\DeclareRobustCommand{\redline}{\raisebox{2pt}{\tikz{\draw[Red,solid,line width = 0.9pt](0,0) -- (5mm,0);}}}

\DeclareRobustCommand{\blackdashedline}{\raisebox{2pt}{\tikz{\draw[-,black!40!black,dashed,line width = 0.9pt](0,0) -- (5mm,0);}}}

\DeclareRobustCommand{\orangedashedline}{\raisebox{2pt}{\tikz{\draw[BurntOrange,dashed,line width = 0.9pt](0,0) -- (5mm,0);}}}

\DeclareRobustCommand{\tikzcircle}[2][red,fill=red]{\tikz[baseline=-0.5ex]\draw[#1,radius=#2] (0,0) circle ;}

\usepackage{multirow}
\usepackage{graphicx}
\usepackage{caption}

\title{Efficient computation of temporal exceeding probability of ship responses in a random wave field}
\author{Xianliang Gong, Katerina Siavelis, Zhou Zhang, Yulin Pan\footnote{yulinpan@umich.edu}}
\date{\small \textit{Department of Naval Architecture and Marine Engineering, University of Michigan, 48109, MI, USA}}

\begin{document}
\maketitle
\section*{Abstract}

In this work, we develop a computational framework to efficiently quantify the temporal exceeding probability of ship responses in a random wave field, i.e., the fraction of time that ship responses exceed a given threshold. In particular, we consider large thresholds so that the response exceedance needs to be treated as rare events. Our computational framework builds on the parameterization of wave field into wave groups and efficient sampling in the group parameter space through Bayesian experimental design (BED). Previous works following this framework exclusively studied extreme statistics of group-maximum response, i.e., the maximum response within a wave group, which is however not straightforward to interpret in practice (e.g., its value depends on the subjective definition of the wave group). In order to adapt the framework to the more robust measure by the temporal exceeding probability, novel developments need to be introduced to the two components of BED (on surrogate model and acquisition function), which are detailed in the paper. We validate our framework for a case of ship roll motion calculated by a nonlinear roll equation, in terms of the agreement of our results to the true solution and the independence of our results to the criterion of defining groups. Finally, we demonstrate the coupling of the framework with CFD models to handle more realistic and general ship response problems.

\vspace{0.2cm}

\noindent \textit{Keywords}: extreme ship responses, temporal exceeding probability, Bayesian experimental design

\section{Introduction}

Extreme ship motions are dangerous events for ships at sea that can be triggered by different mechanisms (e.g., large waves, parametric rolls). Although these large motions occur with a low probability, they may cause severe damage to ships especially at high sea states. With the fast development of high-fidelity CFD models, it is now possible to perform detailed analysis on a single extreme-motion event for a given incoming wave condition. However, the quantification of the statistics of these extreme events still remains a very difficult task.

The difficulty in evaluating the statistics of the extreme motion events lies in the computational cost. Considering the high dimensions (i.e., large number of degrees of freedom) of the wave environment, the rareness of the extreme responses, and the expensiveness of the (high-fidelity) numerical model, a long simulation covering all wave conditions (as the direct Monte Carlo method) is computationally prohibitive. Among many developed approaches to reduce the computational cost (see the review by \cite{belenky2012approaches}), recently a new method based on wave group parameterization and sequential sampling has gained popularity in the marine engineering field \cite{mohamad2018sequential, gong2021full, gong2022sequential, gong2022multi, tang2022estimating}, etc. This new method, as we consider in this paper, first converts the high-dimensional wave field into a low-dimensional parameter space composed by the features of wave groups (in particular the height and length). A sequential sampling method based on the Bayesian experimental design (BED) is then applied to efficiently sample the parameter space (in contrast to a direct Monte-Carlo sampling) which allows the calculation of the statistics of extreme motion events with an affordably small number of samples.

In applying the new method to problems of ship/structure responses, all existing works exclusively focused on the statistics of group-maximum response, i.e., the maximum ship response within a given wave group. While the group-maximum statistics is indeed a natural metric to compute under this methodology (due to the wave group parameterization), this concept is associated with at least two issues for practical interpretation. First, the group-maximum statistics does not accord with the widely-used engineering design metrics, e.g., the return period, and thus may not be intuitive to interpret for ship designers. Second, the group-maximum statistics, say, exceedance probability of group-maximum response, depends on the subjective definition of wave groups, e.g., different definitions lead to different number of groups which affect the denominator in calculating the probability. The second issue becomes more severe with the increase of spectral bandwidth of the wave field, which is accompanied by increasingly unclear structure of wave groups.  Therefore, a single result of the group-maximum probability can be meaningless unless the exact criterion to define wave groups (or the frequency of wave group occurrence) are available to engineers and designers, which is however a cumbersome message to convey. 

In contrast, the temporal exceeding probability (i.e., percentage of the exposure time that ship responses are greater than a given large threshold) provides a robust and universal measure of the extreme event probability that naturally overcomes the ambiguities associated with group-maximum statistics. In computing the temporal exceeding probability, the group parameterization remains as an effective way for dimension reduction of the wave field (although the final result does not depend on the defined groups as we will demonstrate later in the paper). One may expect to further consider the motion exceeding time as the response function of group parameters and apply the sequential sampling enabled by the BED. However, this operation introduces additional difficulties that need to be resolved by substantial developments in both components involved in the BED --- the surrogate model and acquisition function.  While a standard Gaussian process regression (GPR) is sufficient as a surrogate model to compute the group-maximum response (as a smoothly varying response function), great challenges, i.e., prediction errors, arise if the group exceeding time is used as a response function. This is because the group exceeding time as a response function is characterized by two drastically different scales of variation in the regions of zero values (for majority of the inputs) and positive values (for ``dangerous'' critical wave groups) separated by the limiting state.  In addition, the existing acquisition functions for exceeding probability \cite{teixeira2021adaptive, echard2011ak, bichon2008efficient, wang2016gaussian, hu2016global, li2011efficient, sun2017lif, zhu2016reliability} mainly focus on sampling at the limiting state, which becomes insufficient for temporal exceeding probability for which the wave groups leading to longer exceeding time matter more than those at the limiting state.

In this paper, we address the above two problems in the BED procedure, enabling an efficient computation of temporal exceeding probability of ship responses in a random wave field. Specifically, we construct a uniformly-varying (i.e., varying with the same or comparable length scale) response function derived from the group exceeding time which eliminates the prediction errors in GPR while not affecting the final solution of exceeding probability. We then formulate a new acquisition function focusing on sampling wave groups associated with significant exceeding time rather than at the limiting state. We validate our developed computational framework in a case of ship response calculated by a nonlinear roll equation, in terms of the agreement of our obtained result with the true solution, and the independence of our result to different criterion to define wave groups. Finally, we demonstrate the coupling of our framework to CFD models to enable the computation for more realistic and general ship response problems.

The python code for the algorithm, named gpship, is available on Github \footnote{https://github.com/umbrellagong/gpship}.

\section{Method}

\subsection{Problem setup}
\label{problem}

We start from a wave field with a sufficiently long time series of wave elevation $\eta(t), t \in [0, T_{end}]$. Our objective is to compute the temporal exceeding probability (i.e., percentage of time that the ship response is larger than a threshold $r_s$) when a ship goes through the wave field described by time series $\eta(t)$, defined as
\begin{equation}
    P_{temp}  = \frac{\int_{0}^{T_{end}} \mathbf{1}_{|r| > r_s}(r(t)) \mathrm {d} t} {T_{end}},
\label{eq:ptrue}
\end{equation}
where $r(t)$ is the time series of ship response caused by waves $\eta(t)$, and $\mathbf{1}_{|r| > r_s}$ is an indicator function with threshold $r_s$:
\begin{equation}
    \mathbf{1}_{|r| > r_s}(r(t)) = \left\{
    \begin{aligned}
        & 1,         & 
        & if \; |r(t)| > r_s                        \\
        & 0,                             & 
        & if \; |r(t)| \leq r_s  
    \end{aligned}
    \right..
\label{indicator}
\end{equation}
The computation of $P_{temp}$, as defined in \eqref{eq:ptrue}, involves a simulation of the ship response from $\eta(t)$, which needs to be extremely long to cover (many times) all wave conditions associated with a given wave spectrum. If high-fidelity models (e.g., CFD) are used for this computation,  the computational cost can become prohibitively high.

\begin{figure}
    \centering
    \includegraphics[width =14cm]{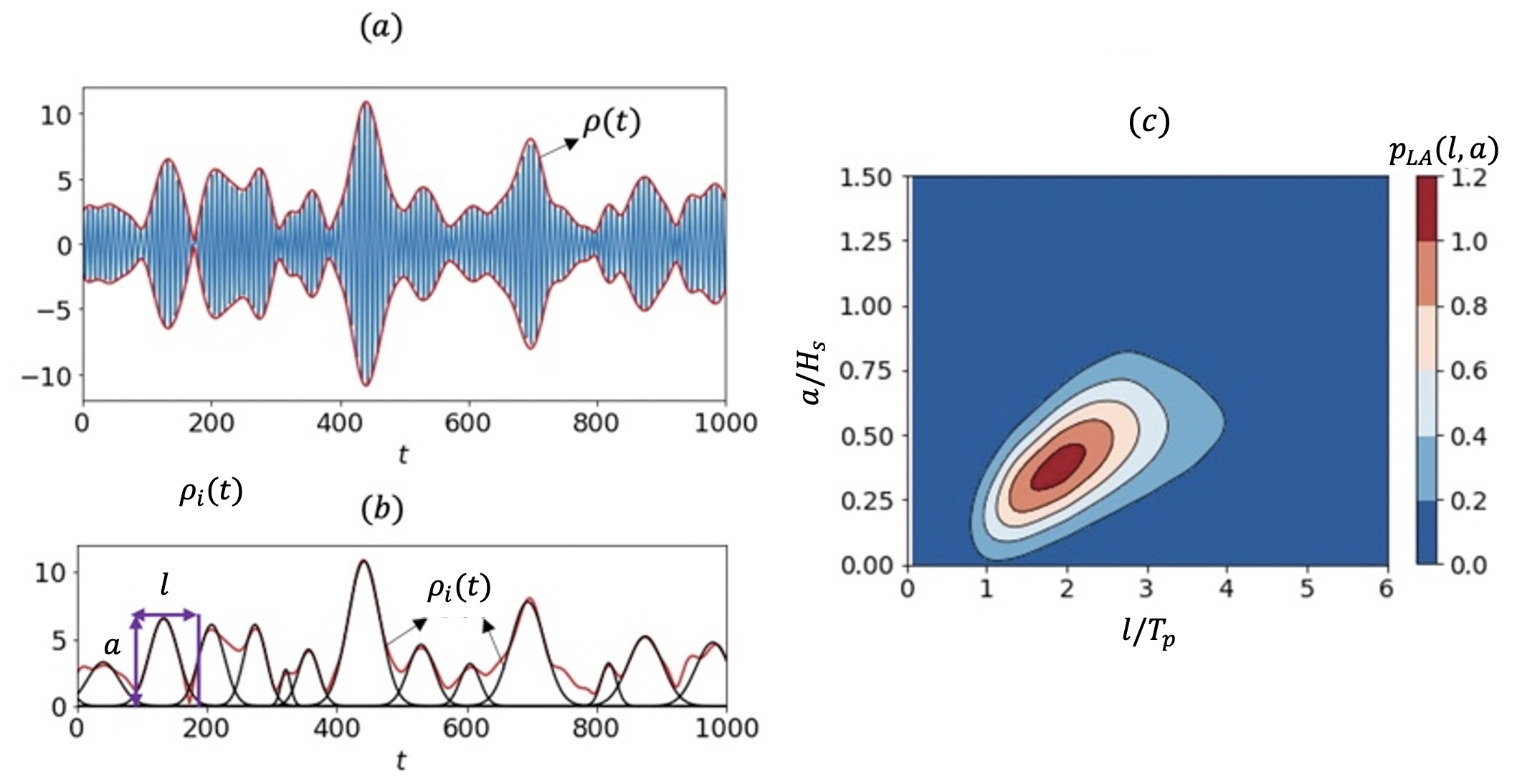}
    \caption{(a) surface elevation $\eta(t)$ (\blueline) and the corresponding envelope process $\rho(t)$ (\redline) in a random wave field. (b) $\rho(t)$ (\redline) fitted by an ensemble of Gaussian wave groups $\rho_i(t)$(\blackline) with parameters $l$ and $a$.
    (c) probability distribution $p_{LA}(l,a)$ obtained from the whole wave field.}
    \label{fig:wavefield}
\end{figure}

In order to reduce the overall computational cost, we can parameterize the time series $\eta(t)$ into wave groups. Specifically, we compute the envelope process $\rho(t)$ from $\eta(t)$ through the Hilbert transform \citep{shum1984estimates} (figure \ref{fig:wavefield}(a)), and then construct Gaussian-like wave groups $\rho_i(t)$ which best fit $\rho(t)$ locally: 
\begin{equation}
    \rho_i(t) = a_i \exp \frac{-(t-t_i)^2}{2l_i^2},
    \label{eq:Gaussian}
\end{equation}
where $t_i$, $a_i$, and $l_i$ are respectively the temporal location, amplitude and length of group $i$ that are selected to fit $\rho(t)$ (see figure \ref{fig:wavefield}(b)). This construction relies on a group detection algorithm proposed by \cite{cousins2016reduced} and later improved in \cite{gong2021full}\footnote{https://github.com/umbrellagong/wavefinder}. The final identification of a wave group is based on the tolerance of a discrepancy index 
\begin{equation}
    D_i(l_i, a_i, t_i) = \frac{\int_{t_i - 2 l_i}^{t_i + 2 l_i}\big(\rho(t) - \rho_i(t) \big)^2 \mathrm{d}t}{\int_{t_i - 2 l_i}^{t_i + 2 l_i} \rho_i(t)^2 \mathrm{d}t}.
    \label{Cmeasure}
\end{equation}
Only groups with $D_i<D^{thr}$ are qualified as wave groups in the time series $\eta(t)$, with $D^{thr}$ serving as a user-defined (subjective) criterion to define wave groups.

Based on the detected groups, we can construct an approximation of $P_{temp}$ as
\begin{equation}
    P_{temp}^a = \frac{\sum_{i=1}^{m}{S(l_i, a_i)}} {T_{end}} \equiv \frac{m}{T_{end}} \int S(l, a)p_{LA}(l,a) \; \mathrm{d}l \mathrm{d}a,
\label{eq:papp}
\end{equation}
where the input-to-output (ItO) function $S(l_i, a_i)= \int \mathbf{1}_{|r| > r_s}(r(t; l_i,a_i))  \mathrm {d} t$ is the time of responses $r(t;l_i,a_i)$ exceeding $r_s$ in group $(l_i,a_i)$, $m$ is the total number of groups in $\eta(t)$. In \eqref{eq:papp}, $P_{temp}^a$ is expressed in two equivalent ways, the first through the summation of exceeding times over all groups in $\eta(t)$, and the second through sampling in the parameter space $(L,A)$ with known probability distribution $p_{LA}(l,a)$ obtained from $\eta(t)$ (see figure \ref{fig:wavefield}(c)). We note that $P_{temp}^a$ is an approximation to $P_{temp}$ (\eqref{eq:papp} relative to \eqref{eq:ptrue}) since certain information is lost when the group parameterization is conducted, including wave phases and the initial condition of the ship encountering a wave group \cite{gong2022snh}. In this work, we focus on the efficient sampling method to compute $P_{temp}^a$, and discuss how the proposed method can be further developed for the computation of $P_{temp}$ in \S3.

We also remark that although the definition in \eqref{eq:papp} involves group numbers $m$ (that depends on the group definition criterion through $D^{thr}$), $P_{temp}^a$ can be considered to be independent of $D^{thr}$ in practical calculation. This can be easily understood from \eqref{eq:papp} especially through the definition with summation of $S$. As long as the large groups leading to positive $S(l_i,a_i)$ are correctly identified, $P_{temp}^a$ remains constant even though the number of small wave groups varies. This is generally true with our group detection algorithm where the threshold $D^{thr}$ only affects the identification of small wave groups (that are ambiguous in nature). One may alternatively argue that another measure could be defined as the number of groups leading to the exceedance within $[0,T_{end}]$, which is also invariant due to the above argument and involves only the group maximum in its computation. However, such measure provides much less information since $S(l_i,a_i)$ additionally captures the ``severity'' of a given wave group by the exceeding time.

Our next task is to design an efficient method for the computation of $P_{temp}^a$ without going through the physical computation of $S(l_i, a_i)$ for each group. In principle this will be achieved through Bayesian experimental design (BED) to compute the  response from a few informative groups for the construction of function $S(l, a)$. In the following sections, we will introduce two basic components of BED specifically in the context of the computation of $P_{temp}^a$: (1) an inexpensive surrogate model to obtain $S$ based on Gaussian process regression (GPR), and (2) an acquisitive function to sequentially select the next-best sample for acceleration of the convergence to $P_{temp}^a$.

\subsection{Surrogate model}
As a general procedure in BED, we may construct a surrogate model for $S(l, a)$ through the Gaussian progress regression (GPR). Given a dataset $\mathcal{D}=\{(l,a)^i, y^i\}_{i=1}^{i=n}$ consisting of $n$ inputs and the corresponding outputs $y=f(l,a)$, our objective is to infer the underlying function $f$, which can be expressed by a Gaussian process 
\begin{equation}
    f(l,a)|\mathcal{D} \sim \mathcal{N}(\mu_{f}(l,a|\mathcal{D}), \sigma^2_{f}(l,a|\mathcal{D})),
\label{eq:post}
\end{equation}
with $\mu_f$ and $\sigma^2_{f}$ respectively the posterior mean and variance. The details of the algorithm and formulae are summarized in Appendix 1.

\begin{figure}
    \centering
    \includegraphics[width=15cm]{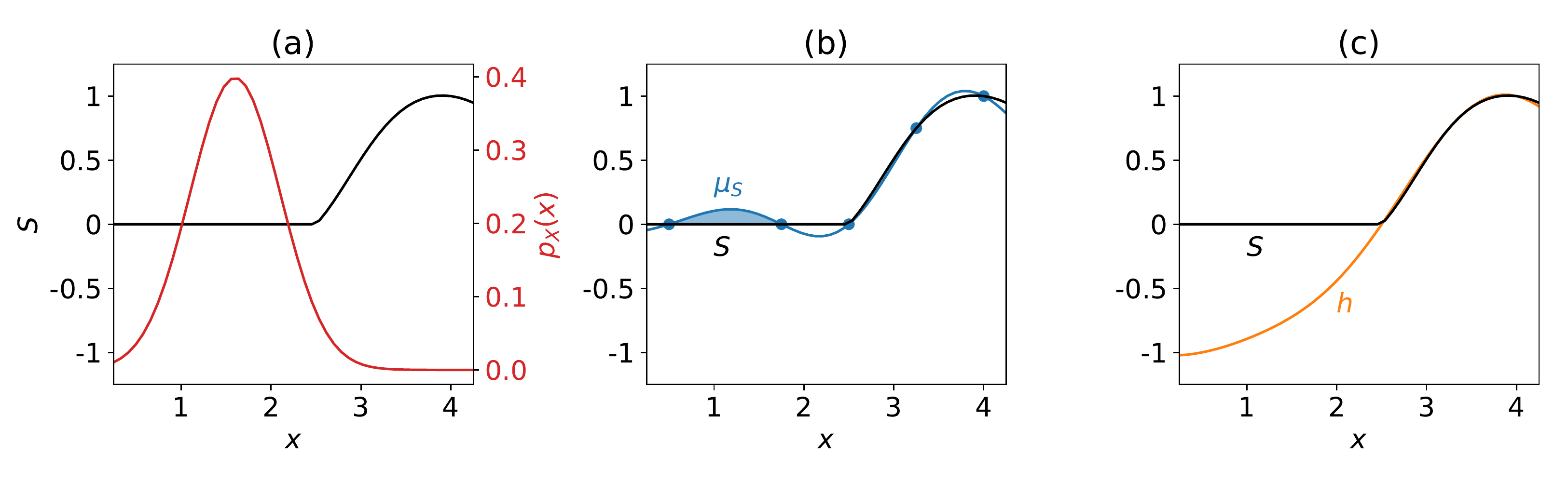}
    \caption{A one-dimensional (1D) demonstration of the issues regarding GPR for exceeding time. This example can be perceived with variable $x$ as the group amplitude $a$ with fixed group length, say $l=1$. (a) The response function of exceeding time $S(x)$ (\blackline) and the input probability distribution $p_X(x)$ (\redline); (b) The function $S(x)$ (\blackline) and the mean of GPR prediction $\mu_S(x)$ (\blueline) with five sample points (\tikzcircle{2pt, Blue}), as well as the false positive region as the shaded area; (c) The function $S(x)$ (\blackline) and the auxiliary function $h(x)$ (\orangeline).}
    \label{fig:penalty}
\end{figure}

However, a direct application of the GPR on function $S(l,a)$ can be problematic, especially in the context of the computation in \eqref{eq:papp}. By definition as the exceeding time, the function $S$ is characterized by zero function values for majority of the input parameters $(l,a)$ and positive for only critical ranges of input. The two regions are associated with drastically different scales of functional variation (the former as a constant with no variation and the latter with much faster variation). In addition, the probability of the input is usually heavily placed on the region where the function value is zero (i.e., majority of groups leading to no exceedance), with the situation illustrated in a phenomenological 1D example in figure \ref{fig:penalty}(a). If a GPR is placed on the dataset from such a function, the predicted function is sketched in \ref{fig:penalty}(b), where prediction errors are inevitably associated with the part with $S=0$. These errors are devastating for the computation of $P_{temp}^a$ when the false positive value of $S$ is accompanied by a large input group probability $p_{LA}(l,a)$, leading to a significant false exceeding probability (considering that the true value of $P_{temp}^a$ is small). This situation can only be alleviated if we have a (very) large number of samples in the region of $S=0$, but this contradicts the goal of the sequential sampling to place the emphasis in regions with large exceeding time. 

To solve this problem, we define an auxiliary function $h(l,a)$ derived from $S(l,a)$, where $h$ contains the information about exceeding time in $S$ and is favorable for GPR to learn. Ideally, $h$ is required to be a function of uniform scale of variation and free of false positive value of exceedance. Considering these constraints, a function $h$ can be defined as
\begin{equation}
h(l,a) = \left\{
\begin{aligned}
    & \frac{S(l,a)}{l},                   & if \; S(l,a) > 0   \\
    & \frac{r_{max}(l,a) - r_s} {r_s},        & if \; S(l,a) = 0 
\end{aligned}
\right..
\end{equation}
where $r_{max}(l,a)$ is the group-maximum response in the group $(l,a)$. When $S=0$, function $h$ takes negative values with $r_{max}-r_s$ serving as a ``negative penalty'' quantifying how far the response is from the threshold. In such a way, the false positive value of $S$ can be avoided. We also normalize the two piecewise segments of $h$ respectively by factors $l$ and $r_s$ so that both segments are in $O(1)$, ensuring the uniform scale of variation. Figure \ref{fig:penalty}(c) demonstrates a typical function $h$ in the 1D example.

After the GPR of $h$ is available, we can recover function $S$ by $S \equiv \mathbf{1}_{h>0}(h) \, h \,l$. Such recovered $S$ is also free of false positive values because of the ``negative penalty'' placed on $h$.

\subsection{Acquisition function}

Given the function $h$ (and $S$) learned from the GPR based on dataset $\mathcal{D}$, $P_{temp}^{a}$ can be considered as a random variable with its randomness resulting from the uncertainty in $h$. The uncertainty in $P_{temp}^{a}$ can therefore be estimated as
\begin{align}
    U(\mathcal{D}) 
    & =  \frac{\sum_{i=1}^{m}  \big|S^+(l_i,a_i|\mathcal{D}) - S^-(l_i,a_i|\mathcal{D})\big| }{T_{end}},
\label{eq:u1}
\end{align}
where 
$S^{\pm}(l_i,a_i|\mathcal{D})=\mathbf{1}_{h>0}\big(h^{\pm}(l_i,a_i|\mathcal{D})\big) \, h^\pm(l_i,a_i|\mathcal{D}) \, l_i$, with $h^{\pm}$ the upper and lower bound of $h$ estimated by (one standard deviation up and below)
\begin{equation}
    h^{\pm}(l,a|\mathcal{D}) = \mu_h (l,a|\mathcal{D}) \; {\pm} \; \sigma_h (l,a|\mathcal{D}).
\label{eq:g+}
\end{equation}

Our purpose is to select the next sample, after adding which the uncertainty in $P_{temp}^{a}$ is significantly reduced. For an efficient way to fulfill this purpose, we further formulate the uncertainty in $P_{temp}^{a}$ after adding one hypothetical sample at $\tilde{l},\tilde{a}$:
\begin{align}
    U(\mathcal{D}, \tilde{l},\tilde{a})  = &   \frac{\sum_{i=1}^{m}  \big|S^+\big(l_i,a_i|\mathcal{D},\overline{h}(\tilde{l},\tilde{a})\big) - S^-\big(l_i,a_i|\mathcal{D},\overline{h}(\tilde{l},\tilde{a})\big) \big| }{T_{end}} 
\label{eq:u2}
\end{align}
with $\overline{h}(\tilde{l},\tilde{a}) = \mu_{h}(\tilde{l},\tilde{a}|\mathcal{D})$ the mean prediction in \eqref{eq:post}. The computation of $S^{\pm}\big(l_i,a_i|\mathcal{D},\overline{h}(\tilde{l},\tilde{a})\big)$ relies on $h^{\pm}(l_i,a_i| \mathcal{D}, \overline{h}(\tilde{l},\tilde{a}))$, with the formulation of the latter detailed in Appendix \ref{Appendix1}. 


The selection of next sample can then be formulated as an optimization problem to minimize the hypothetical next-step uncertainty:
\begin{equation}
    l^*,a^* = {\rm{argmin}}_{\tilde{l},\tilde{a}}  U(\mathcal{D}, \tilde{l},\tilde{a}),
\label{eq:opt}
\end{equation}
which can be solved using standard optimization methods. In our work, we apply a combined brute-force grid search (with coarse grid) and a gradient-based (for fine search) method\footnote{\url{https://github.com/scipy/scipy/blob/v1.8.1/scipy/optimize/_optimize.py#L3300-L3554}} in the two-dimensional space.

\begin{figure}
    \centering
    \includegraphics[width=16cm]{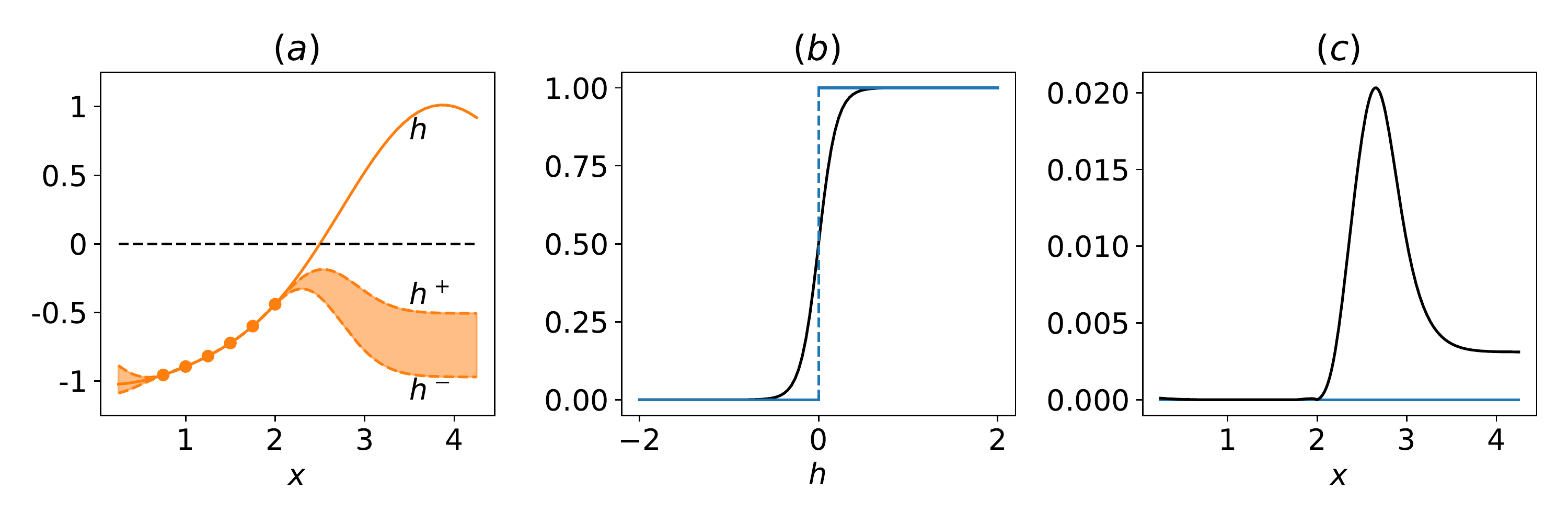}
    \caption{A 1D demonstration of the effect of softer indicator function. This example can be perceived with variable $x$ as the group amplitude $a$ with fixed group length, say $l=1$. (a) upper and lower bounds $h^\pm(x)$ (\orangedashedline) generated from sample points on true function $h(x)$ (\orangeline) that are far from exceedance; (b) The original indicator function $\mathbf{1}_{h>0}(h)$ (\blueline) and the soft indicator function $1 / (1+e^{-c \, h})$ with $c=10$ (\blackline); (c) The uncertainty $|S^+(x|\mathcal{D}) - S^-(x|\mathcal{D})|$ as a function of $x$, computed by the original (\blueline) and soft (\blackline) indicator functions.}
    \label{fig:logistic}
\end{figure}

We note that if the (initial) dataset $\mathcal{D}$ contains only samples far from exceedance, $h^{\pm}(l_i, a_i)$ can be negative for all wave groups (see figure \ref{fig:logistic}(a)). As a result,  $S^{\pm}(l_i, a_i)=0$ and the uncertainty defined in \eqref{eq:u2} vanishes (figure \ref{fig:logistic}(c)). Under these situations, in order to robustly initiate the sequential sampling, one can temporarily apply a `soft' indicator function in computing $S^{\pm}(l_i, a_i)$, i.e., we use a logistic function  $1 / (1+e^{-c \, h^\pm})$ with $c \gg 1$ instead of $\mathbf{1}_{h>0}(h^\pm)$ (figure \ref{fig:logistic}(b)). This procedure replaces the zero values of $S^{\pm}(l_i, a_i)$ by small positive values, but with meaningful uncertainties (represented by the upper and lower bounds) to drive the next sample to regions with larger exceedance (figure \ref{fig:logistic}(c)).

We summarize the full BED algorithm in Algorithm 1.

\begin{algorithm}
    \caption{BED for temporal exceeding probability}
  \begin{algorithmic}
    \REQUIRE Number of initial samples $n_{init}$ and sequential samples $n_{seq}$
    \INPUT Initial dataset $\mathcal{D}=\{(l^i,a^i), h^i\}_{i=1}^{n_{init}}$
    \STATE \textbf{Initialization} $j = 0$
    \WHILE{$j < n_{seq}$}
      \State Train the surrogate model \eqref{eq:post} with $\mathcal{D}$
      \STATE Solve \eqref{eq:opt} to find the next best sample $(l^{j+1},a^{j+1})$
      \STATE Implement numerical simulation to get $ h^{j+1} = h(l^{j+1},a^{j+1})$
      \STATE Update the dataset  $\mathcal{D} = \mathcal{D} \cup \{(l^{j+1}, a^{j+1}), h^{j+1}\}$
      \STATE $j = j + 1$
    \ENDWHILE
\OUTPUT Compute the temporal exceeding probability \eqref{eq:papp} based on the  surrogate model
  \end{algorithmic}
\label{al}
\end{algorithm}

\section{Validation of the Method}
In this section, we compute $P^a_{temp}$ by the proposed sequential sampling approach, with comparison to the result from the space-filling Latin hypercube (LH) sampling and the true solution of $P^a_{temp}$. In addition, we will discuss the invariance of $P^a_{temp}$, the difference between $P^a_{temp}$ and $P_{temp}$ and suggest improved methods to eventually capture $P_{temp}$. Since $P^a_{temp}$ and $P_{temp}$ need to be accurately evaluated, an efficient ship response simulator is required. For this work, we  use a nonlinear roll equation to calculate $r(t)$ from a given wave signal $\eta(t)$:
\begin{equation}
    \ddot r + \alpha_1\dot r + \alpha_2\dot r^3 + (\beta_1+ \epsilon_1 \cos(\theta)\eta(t)) r+\beta_2 r^3=\epsilon_2 \sin(\theta)\eta(t),
\label{eq:roll}
\end{equation}
which models the ship roll response due to nonlinear resonance and parametric roll in oblique irregular waves. Empirical coefficients are set as $\alpha_1 = 0.35$, $\alpha_2 = 0.06$, $\beta_1 = 0.04$, $\beta_2 = -0.1$, $\theta=\pi / 6$, $\epsilon_1 = 0.016$, and $\epsilon_2 = 0.012$. The wave field to be decomposed into groups is generated from a narrow-band spectrum of a Gaussian form:
\begin{equation}
F(\omega) =  \frac{H_s^2}{16}\frac{1}{\sqrt{2\pi}d}\exp (\frac{-(\omega-\omega_p)^2}{2 d^2}),  
\label{Fk}
\end{equation}
with $\omega$ the angular frequency, $H_s=12m$ the significant wave height , $\omega_p=0.067 s^{-1}$ the peak (carrier) wave frequency (corresponding to peak period $T_p=15s$), and $d=0.02 s^{-1}$ a parameter of the spectral bandwidth. 

Following procedures in \S 2.1, the wave field described by \eqref{Fk} can be reduced to a parameter space $(L,A)$ with known probability $p_{LA}(l,a)$ (cf. figure \ref{fig:wavefield}(c)). In computing $P^a_{temp}$, the response $r(t; l, a)$ from a group $(l,a)$ is needed, which is calculated by simulation of \eqref{eq:roll} in $t \in \{-3l, \;3l\}$ with input $\eta(t;l,a) = a \exp(-\frac{t^2}{2l^2}) \cos(\omega_p t)$ and $(0, 0)$ initial condition at $t=-3l$. We note that the choice of $3l$ does not appreciably affect the final solution as long as the value is sufficiently large to cover the portion of the group with significant amplitude. The exceeding time in one group is then computed by $S(l,a)= \int_{-3l}^{3l} \mathbf{1}_{|r| > r_s}(r(t; l,a))  \; \mathrm {d} t$, and the algorithm described in \S 2 can be applied accordingly.

\begin{figure}[htbp]
\centering
\subfigure[]{
\includegraphics[width=0.46\textwidth]{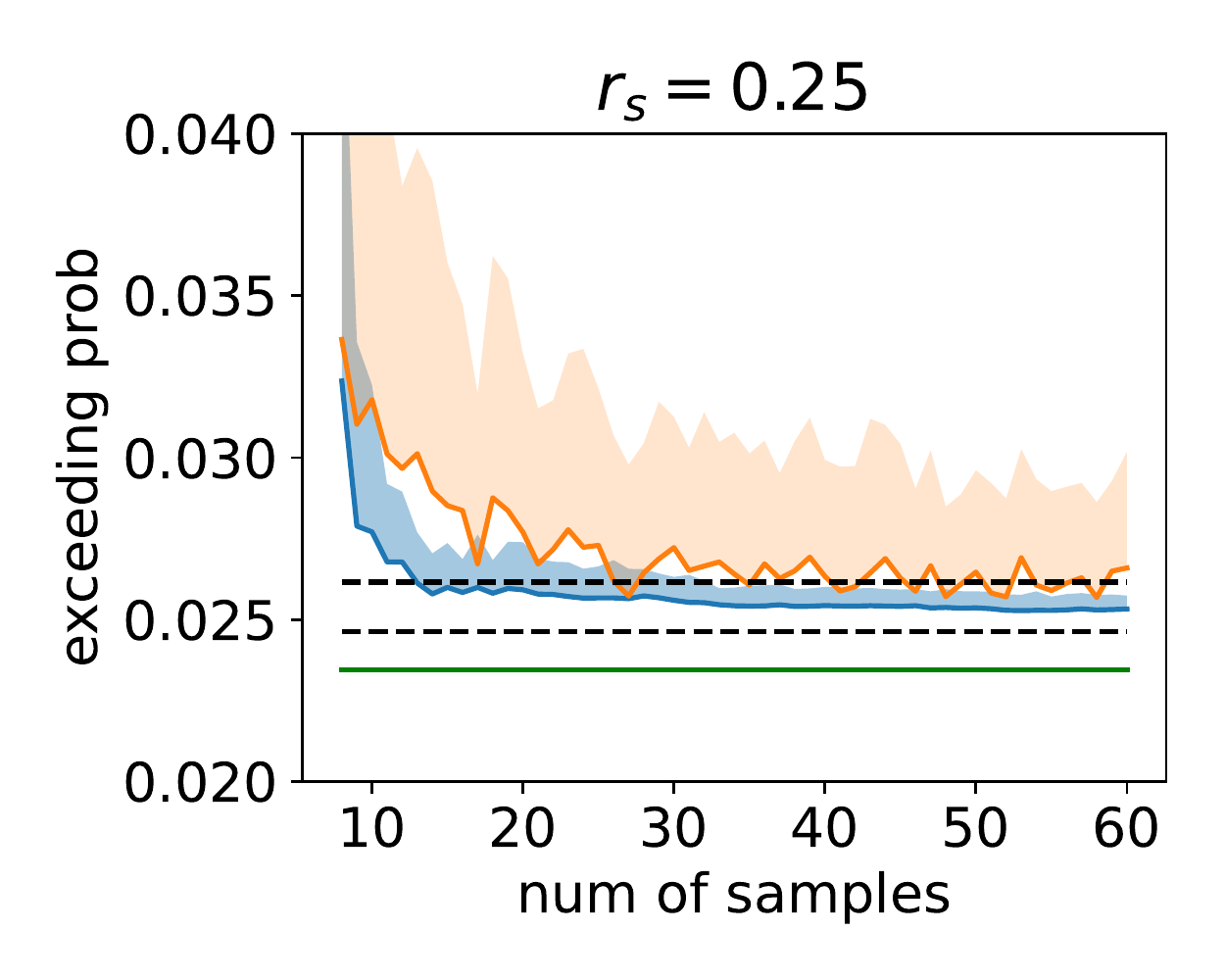}
}
\quad
\subfigure[]{
\includegraphics[width=0.46\textwidth]{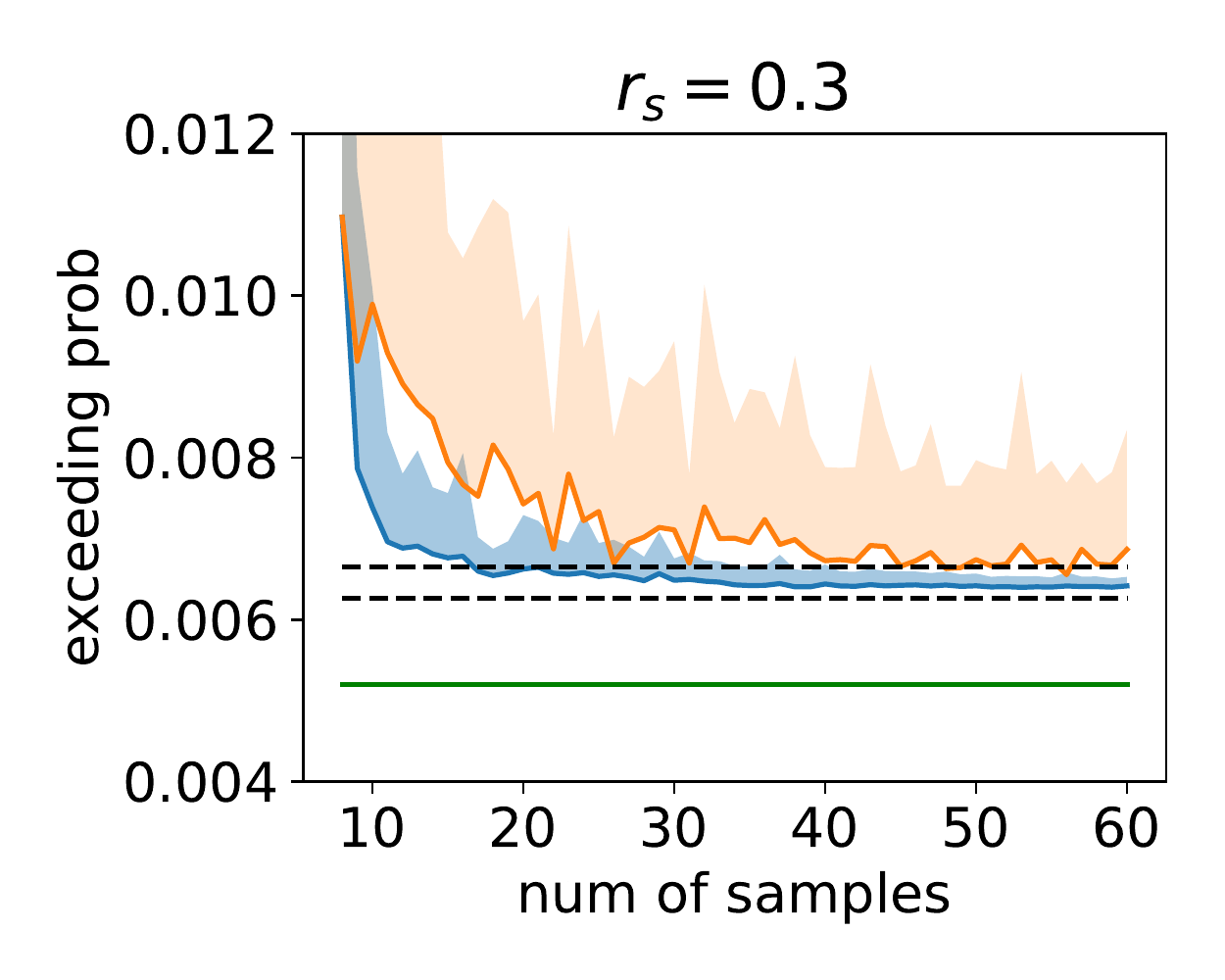}
}
\caption{Temporal exceeding probability $P_{temp}^a$ as a function of sampling numbers, calculated from sequential sampling (\blueline) and LH sampling (\orangeline), for (a) $r_s=0.25$ radians (b) $r_s=0.3$ radians. The shaded region represents one standard deviation above the mean estimated from 100 applications of the corresponding methods. The true solution of $P_{temp}^a$ is shown (\blackdashedline) in terms of the 3\% error bounds, and the true solution of $P_{temp}$ is indicated (\greenline).}
\label{fig:validation1}
\end{figure}

\begin{figure}
    \centering
    \includegraphics[width =8cm]{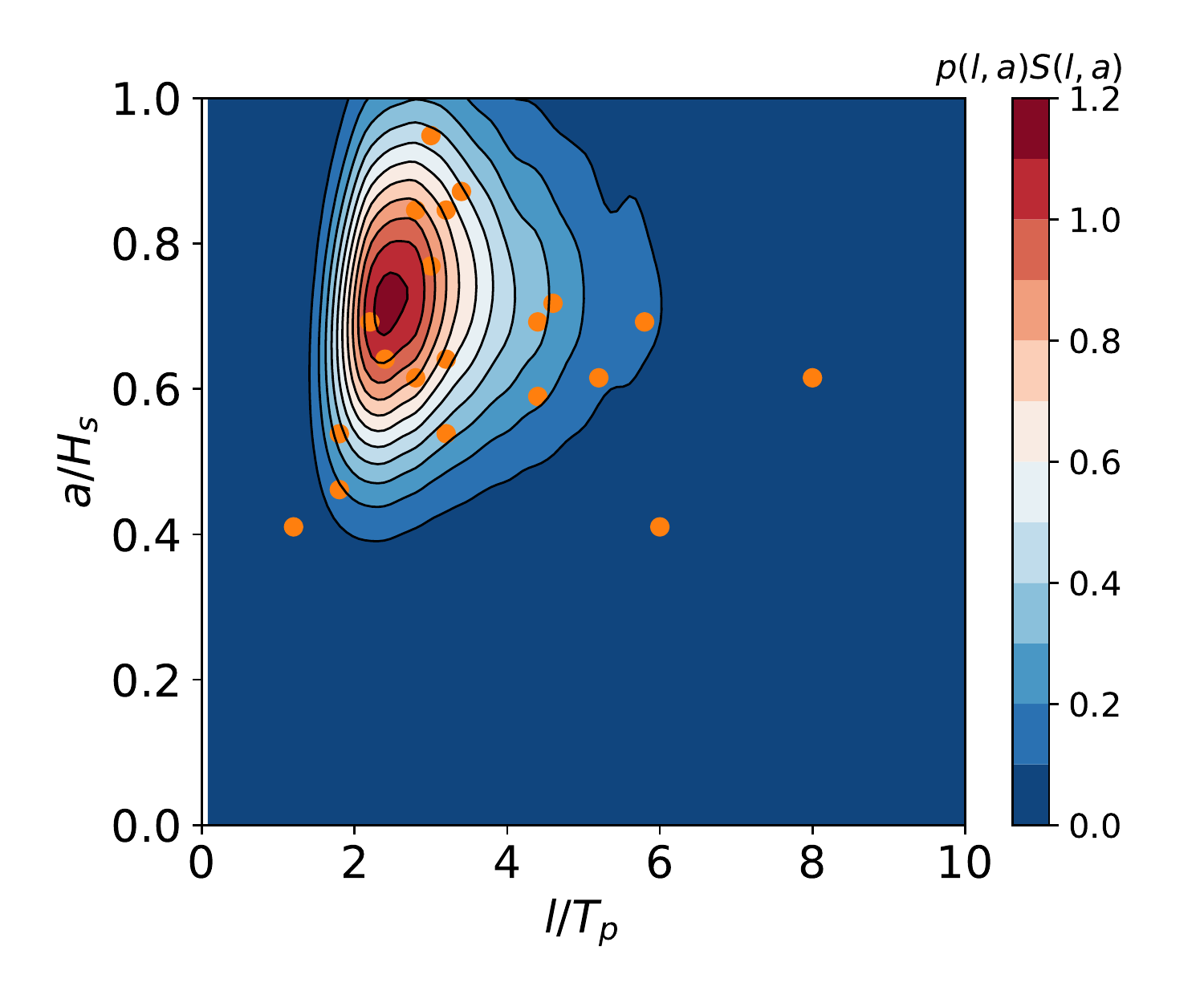}
    \caption{Positions of sequential samples (\tikzcircle{2pt, orange}) in the space of $S(l,a) p_{LA}(l,a)$ with a contour plot.}
    \label{fig:validation2}
\end{figure}

The results of $P^a_{temp}$ for two threshold values $r_s$=0.25 and 0.3 radians are shown in figure \ref{fig:validation1} (a) and (b). In both cases, we present results computed by our new sequential BED sampling and standard LH sampling, along with the true solution of $P^a_{temp}$ and $P_{temp}$ (by brute-force calculations with a large number of groups and a long time series, with the difference in their values discussed at the end of this section). The sequential samplings are conducted with an initial data set of 8 LH samples. Also included in figure \ref{fig:validation1} are the standard deviations of the sequential-sampling and LH-sampling results, with both obtained from 100 applications of the corresponding methods from different initial samples. It is clear that for both values of $r_s$, the sequential-sampling result approaches to the true solution of $P^a_{temp}$ much faster than that by LH sampling, with the former also exhibiting a much smaller standard deviation. In particular, for both cases, it takes $O(20)$ sequential samples for the prediction of $P^a_{temp}$ (including the uncertainty bounds) to fall into the 3\% error range of the true solution. In contrast, the LH-sampling provides a result with much larger uncertainty even at the end of 60 samples, indicating that a single experiment has a large chance to provide a solution that significantly deviates from the true solution of $P^a_{temp}$. 

We further examine the reason for the fast convergence of the sequential-sampling results by plotting the sample positions in the space of $S(l,a)p_{LA}(l,a)$ in figure \ref{fig:validation2}. Here the contour of $S(l,a)p_{LA}(l,a)$ provides a measure of the importance of a group $(l,a)$ in computing $P^a_{temp}$, which can also be seen from \eqref{eq:papp}. As shown in the figure, most sequential samples are driven to the region with significant $S(l,a)p_{LA}(l,a)$, indicating the effectiveness of our BED method with new developments in both the surrogate model and acquisition function.

We next demonstrate the invariance of temporal exceeding probability with group detection criterion, particularly the threshold $D_{thr}$ for \eqref{Cmeasure}. Table 1 lists the values of $P_{temp}^a$ and the group-maximum exceeding probability ($P_{group}^a = \sum_{i=1}^m \mathbf{1}_{r_{max}>r_s}\big(r_{max}(l_i,a_i) \big) / m$) for different $r_s$ and $D_{thr}$. For all values of $r_s$, $P_{group}^a$ changes more than $30\%$ for $D_{thr}$ varying from 0.65 to 0.8. In contrast, $P_{temp}^a$ remains almost a constant, with the very small variation resulting from some small/deformed groups leading to large motion that escape from the detection.

We finally discuss the difference between $P_{temp}^a$ and $P_{temp}$ shown in figure \ref{fig:validation1}. For these cases, $P_{temp}^a$ represents a $O(20\%)$ over-estimation compared to $P_{temp}$. This situation is usually acceptable in practice considering the small value of the exceeding probability itself. However, for general cases, the over-estimation cannot be guaranteed and it is desirable to develop more sophisticated method to directly and efficiently compute $P_{temp}$. We believe that this can be achieved through the combination of the method in this paper and techniques compensating the lost information in the wave group representation by heteroscedastic Gaussian process developed by us in \cite{gong2022sequential}.

\begin{table}[]
\begin{center}
\caption{Comparison of $P_{temp}^{a}$ and $P_{group}^{a}$ for varying $D_{thr}$ and $r_s$.}
\begin{threeparttable}[t]
\begin{tabular}{|c|ccc|ccc|}
\hline
\multirow{2}{*}{$D_{thr}$} &
  \multicolumn{3}{c|}{$P_{temp}^{a}$} &
  \multicolumn{3}{c|}{$P_{group}^{a}$} \\ \cline{2-7} 
 &
  \multicolumn{1}{c|}{$r_s = 0.25$} &
  \multicolumn{1}{c|}{$r_s = 0.3$} &
  $r_s = 0.35$ &
  \multicolumn{1}{c|}{$r_s = 0.25$} &
  \multicolumn{1}{c|}{$r_s = 0.3$} &
  $r_s = 0.35$ \\ \hline
0.35 &
  \multicolumn{1}{c|}{0.02566} &
  \multicolumn{1}{c|}{0.006473} &
  0.0004861 &
  \multicolumn{1}{c|}{0.2632} &
  \multicolumn{1}{c|}{0.09884} &
  0.01335 \\ \hline
0.30 &
  \multicolumn{1}{c|}{0.02564} &
  \multicolumn{1}{c|}{0.006471} &
  0.0004860 &
  \multicolumn{1}{c|}{0.2781} &
  \multicolumn{1}{c|}{0.1047} &
  0.01414 \\ \hline
0.25 &
  \multicolumn{1}{c|}{0.02554} &
  \multicolumn{1}{c|}{0.006462} &
  0.0004859 &
  \multicolumn{1}{c|}{0.3278} &
  \multicolumn{1}{c|}{0.1246} &
  0.01688 \\ \hline
0.20 &
  \multicolumn{1}{c|}{0.02467} &
  \multicolumn{1}{c|}{0.006352} &
  0.0004825 &
  \multicolumn{1}{c|}{0.4050} &
  \multicolumn{1}{c|}{0.1598} &
  0.02206 \\ \hline
variation \tnote{1}  &
  \multicolumn{1}{c|}{3.8 \%} &
  \multicolumn{1}{c|}{1.8 \%} &
  0.7 \% &
  \multicolumn{1}{c|}{35 \%} &
  \multicolumn{1}{c|}{38 \%} &
  39 \% \\ \hline
\end{tabular}
\begin{tablenotes}
     \item[1] The variation is computed by the relative difference (normalized by the largest values) in each column.
\end{tablenotes}
\end{threeparttable}
\end{center}
\end{table}

\section{Coupling to CFD}
\begin{figure}
    \centering
    \includegraphics[width=0.8\textwidth]{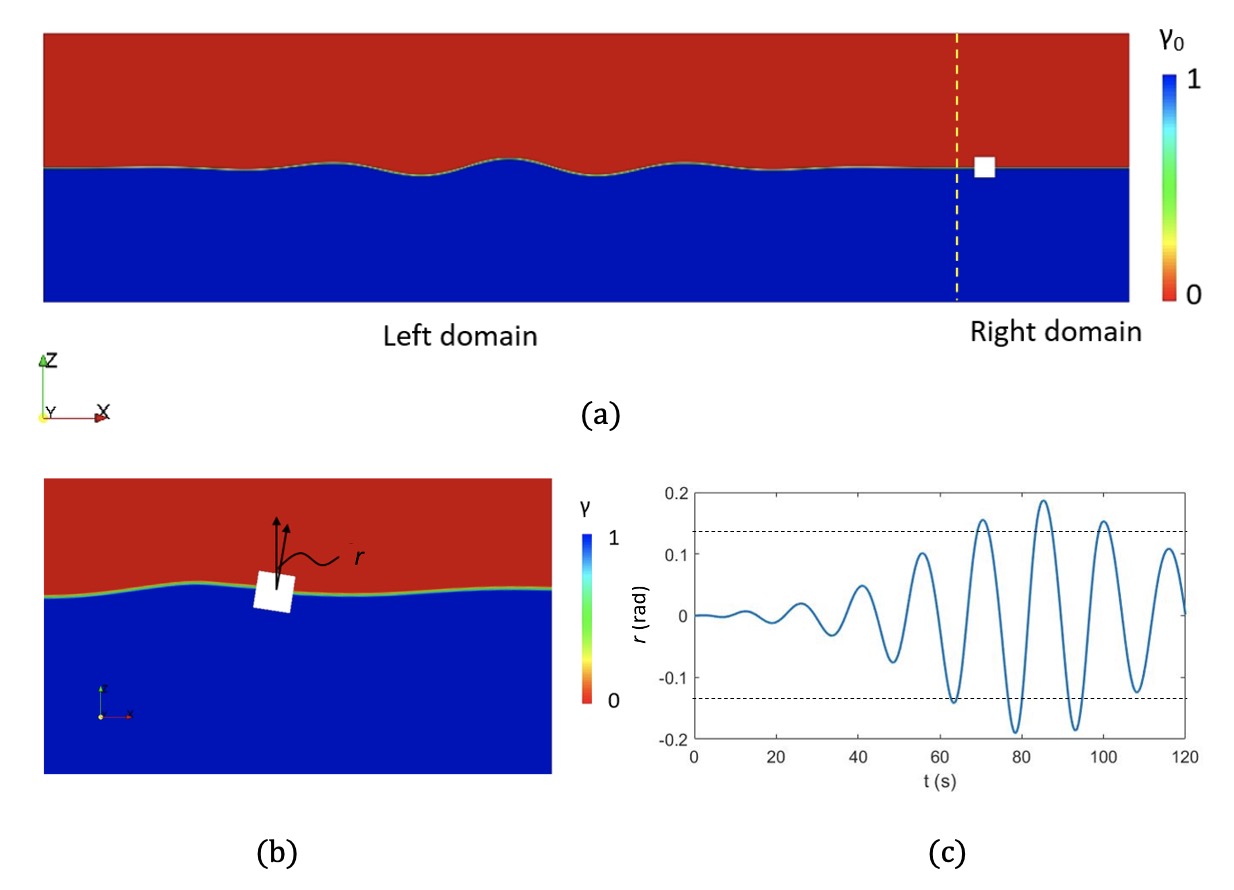}
    \caption{A typical CFD simulation for a wave group with $a = 8.1m$ and $l = 332.7m$. (a) initial wave field with volume fraction $\gamma_0$, with hull located on the right; (b) volume fraction $\gamma$ in the process of a wave group interacting with the hull; (c) time series of $r(t)$ (\blueline) with threshold $r_s$ (\blackdashedline).}
\label{fig:cfd}
\end{figure}

While our sequential BED method has been validated using a low-fidelity ship motion model in the previous section, it may be desired to couple the approach to higher-fidelity (e.g., CFD) models in practical applications. In this section, we demonstrate such an application with CFD simulations to compute the ship roll responses. For simplicity, we consider the motion of a two-dimensional (2D),  square-shaped hull geometry with $40m\times40m$ cross section and density $\rho_h=0.5\rho_w$ with $\rho_w$ being the water density. The group parameters $(l,a)$ are transferred to the CFD simulations as the initial condition (i.e., to simulate an $(l,a)$ group propagating toward the ship hull, see figure \ref{fig:cfd}(a)). The exceeding time are computed from the time series of the simulated roll motion (see figure \ref{fig:cfd}(b) and (c)).

The CFD model is developed using the open-source code OpenFOAM \cite{Jasak_2009}. The interFoam solver is used to capture the air-water interface through an algebraic volume of fluid (AVOF) method. A standard $k-\epsilon$ turbulence model is applied in conjunction with the AVOF method \cite{Mirjalili_2017}. The 2D hull is considered as a rigid body, moving under the force exerted by flow pressure and shear stress. The motion of the hull is calculated by numerical integration implemented by the Newmark method \cite{Newmark_1959}. More details about the boundary conditions, model equations and grid resolutions of the solver can be found in \cite{gong2021full}.

\begin{figure}
    \centering
    \includegraphics[width=12cm]{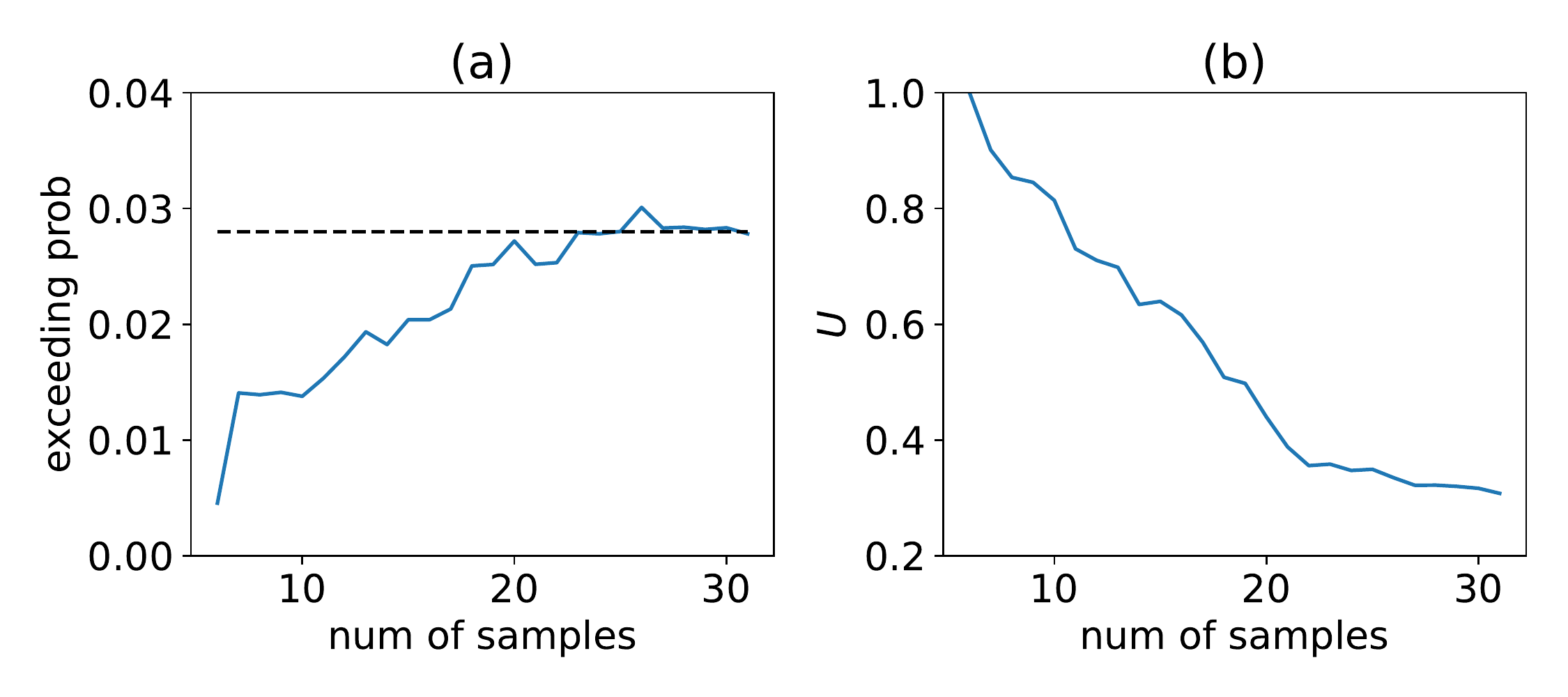}
    \caption{(a) Temporal exceeding probability $P^{a}_{temp}$ (\blueline) for $r_s=0.13$ as a function of the sequential sample number, with the convergent level indicated (\blackdashedline); (b) Normalized uncertainty as a function of the sequential sample number.}
    \label{fig:cfd_2}
\end{figure}

In computing $P_{temp}^a$, we set a threshold of $r_s=0.13$ radians in a wave field of $H_s=9m$ and otherwise the same as that in \S 3. Figure \ref{fig:cfd_2}(a) shows the results of $P_{temp}^{a}$ using 25 sequential samples followed by 6 LH samples. It is clear that the estimated $P_{temp}^{a}$ approaches a constant level after $O(20)$ sequential samples, indicating the convergence of the result. Since the true solution of $P_{temp}^{a}$ is not available for this case (unless much more computational resources can be allocated to run the CFD model, which is not feasible now), we examine the uncertainties \eqref{eq:u1} associated with the solution, normalized by its value at initial time. As shown in figure \ref{fig:cfd_2}(b), the uncertainty decreases rapidly as $P_{temp}^{a}$ approaches a constant level, demonstrating the effectiveness of our method when coupled to CFD.

\section{Conclusion}
In this work, we develop a computational framework to efficiently compute the temporal exceeding probability of ship responses in a random wave field, i.e. the fraction of time that the response exceeds a specified threshold. Compared with group-maximum exceeding probability, temporal exceeding probability provides a more robust measure of the extreme motion due to its invariance regarding different group definitions and consistency with design metrics of return period. To enable the computation of temporal exceeding probability, we develop a novel BED framework incorporating (1) a uniformly-varying response function resulted from negative penalty and normalization of the group exceeding time; (2) a new acquisition function focusing on sampling wave groups associated with significant exceeding time. We validate our framework in the context of a nonlinear roll equation in terms of the efficiency of the sequential sampling and the invariance of results to wave group definitions. We also demonstrate the coupling of our framework to CFD simulations to show its applicability to higher-fidelity models. Future development will focus on heteroscedastic surrogate model \cite{gong2022sequential} and analytical acquisition formula \cite{gong2022multi, blanchard2021output} to enable efficient sampling in high-dimensional parameter space in the context of temporal exceeding probability.

\section*{ACKNOWLEDGEMENT}

This research is supported by the Office of Naval Research
grant N00014-20-1-2096. We thank the program manager Dr. Woei-Min Lin for several helpful discussions
on the research. 
This work used the Extreme Science and Engineering Discovery Environment (XSEDE) through allocation TG-BCS190007.

\bibliographystyle{unsrt}
\bibliography{reference.bib}

\begin{thebibliography}{10}

\bibitem{belenky2012approaches}
Vadim Belenky, Kenneth~M Weems, Christopher~C Bassler, Martin~J Dipper,
  Bradley~L Campbell, and Kostas~J Spyrou.
\newblock Approaches to rare events in stochastic dynamics of ships.
\newblock {\em Probabilistic Engineering Mechanics}, 28:30--38, 2012.

\bibitem{mohamad2018sequential}
Mustafa~A Mohamad and Themistoklis~P Sapsis.
\newblock Sequential sampling strategy for extreme event statistics in
  nonlinear dynamical systems.
\newblock {\em Proceedings of the National Academy of Sciences},
  115(44):11138--11143, 2018.

\bibitem{gong2021full}
Xianliang Gong, Zhou Zhang, Kevin~J Maki, and Yulin Pan.
\newblock Full resolution of extreme ship response statistics.
\newblock {\em arXiv preprint arXiv:2108.03636}, 2021.

\bibitem{gong2022sequential}
Xianliang Gong and Yulin Pan.
\newblock Sequential bayesian experimental design for estimation of
  extreme-event probability in stochastic input-to-response systems.
\newblock {\em Computer Methods in Applied Mechanics and Engineering},
  395:114979, 2022.

\bibitem{gong2022multi}
Xianliang Gong and Yulin Pan.
\newblock Multi-fidelity bayesian experimental design to quantify extreme-event
  statistics.
\newblock {\em arXiv preprint arXiv:2201.00222}, 2022.

\bibitem{tang2022estimating}
Tianning Tang and Thomas~AA Adcock.
\newblock Estimating space--time wave statistics using a sequential sampling
  method and gaussian process regression.
\newblock {\em Applied Ocean Research}, 122:103127, 2022.

\bibitem{teixeira2021adaptive}
Rui Teixeira, Maria Nogal, and Alan O’Connor.
\newblock Adaptive approaches in metamodel-based reliability analysis: A
  review.
\newblock {\em Structural Safety}, 89:102019, 2021.

\bibitem{echard2011ak}
Benjamin Echard, Nicolas Gayton, and Maurice Lemaire.
\newblock Ak-mcs: an active learning reliability method combining kriging and
  monte carlo simulation.
\newblock {\em Structural Safety}, 33(2):145--154, 2011.

\bibitem{bichon2008efficient}
Barron~J Bichon, Michael~S Eldred, Laura~Painton Swiler, Sandaran Mahadevan,
  and John~M McFarland.
\newblock Efficient global reliability analysis for nonlinear implicit
  performance functions.
\newblock {\em AIAA journal}, 46(10):2459--2468, 2008.

\bibitem{wang2016gaussian}
Hongqiao Wang, Guang Lin, and Jinglai Li.
\newblock Gaussian process surrogates for failure detection: A {Bayesian}
  experimental design approach.
\newblock {\em Journal of Computational Physics}, 313:247--259, 2016.

\bibitem{hu2016global}
Zhen Hu and Sankaran Mahadevan.
\newblock Global sensitivity analysis-enhanced surrogate (gsas) modeling for
  reliability analysis.
\newblock {\em Structural and Multidisciplinary Optimization}, 53(3):501--521,
  2016.

\bibitem{li2011efficient}
Jing Li, Jinglai Li, and Dongbin Xiu.
\newblock An efficient surrogate-based method for computing rare failure
  probability.
\newblock {\em Journal of Computational Physics}, 230(24):8683--8697, 2011.

\bibitem{sun2017lif}
Zhili Sun, Jian Wang, Rui Li, and Cao Tong.
\newblock Lif: A new kriging based learning function and its application to
  structural reliability analysis.
\newblock {\em Reliability Engineering \& System Safety}, 157:152--165, 2017.

\bibitem{zhu2016reliability}
Zhifu Zhu and Xiaoping Du.
\newblock Reliability analysis with monte carlo simulation and dependent
  kriging predictions.
\newblock {\em Journal of Mechanical Design}, 138(12), 2016.

\bibitem{shum1984estimates}
KT~Shum and W~Kendall Melville.
\newblock Estimates of the joint statistics of amplitudes and periods of ocean
  waves using an integral transform technique.
\newblock {\em Journal of Geophysical Research: Oceans}, 89(C4):6467--6476,
  1984.

\bibitem{cousins2016reduced}
Will Cousins and Themistoklis~P Sapsis.
\newblock Reduced-order precursors of rare events in unidirectional nonlinear
  water waves.
\newblock {\em Journal of Fluid Mechanics}, 790:368--388, 2016.

\bibitem{gong2022snh}
Xianliang Gong and Yulin Pan.
\newblock Effects of varying initial conditions of ship encountering wave
  groups in computing extreme motion statistics.
\newblock {\em 34rd Symposium on Naval Hydrodynamics}, 2022.

\bibitem{Jasak_2009}
Hrvoje Jasak.
\newblock Open{FOAM}: Open source {CFD} in research and industry.
\newblock {\em International Journal of Naval Architecture and Ocean
  Engineering}, 1(2):89–94, 2009.

\bibitem{Mirjalili_2017}
S.~Mirjalili, S.~S. Jain, and M.~Dodd.
\newblock Interface-capturing methods for two-phase flows: An overview and
  recent developments.
\newblock {\em Annual Research Briefs}, page 117–135, 2017.

\bibitem{Newmark_1959}
Nathan~M. Newmark.
\newblock A method of computation for structural dynamics.
\newblock {\em Journal of the Engineering Mechanics Division}, 85:67--94, 1959.

\bibitem{blanchard2021output}
Antoine Blanchard and Themistoklis Sapsis.
\newblock Output-weighted optimal sampling for bayesian experimental design and
  uncertainty quantification.
\newblock {\em SIAM/ASA Journal on Uncertainty Quantification}, 9(2):564--592,
  2021.

\bibitem{rasmussen2003gaussian}
Carl~Edward Rasmussen.
\newblock Gaussian processes in machine learning.
\newblock In {\em Summer School on Machine Learning}, pages 63--71. Springer,
  2003.

\end{thebibliography}

\appendix

\section{Gaussian process regression}
\label{Appendix1}
In this section, we briefly outline the Gaussian process regression \cite{rasmussen2003gaussian}. Consider the task of inferring $f$ from $\mathcal{D}=\{\boldsymbol{x}^i, y^i\}_{i=1}^{i=n}$ consisting of $n$ inputs $\mathbb{X} = \{\boldsymbol{x}^{i}\in \mathbb{R}^d \}_{i=1}^{i=n}$ and the corresponding outputs $\boldsymbol{y} = \{y^{i}\in \mathbb{R}\}_{i=1}^{i=n}$ (where $\boldsymbol{x}=(l,a)$ and $f=h$ in our problem). In GPR, a prior, representing our beliefs over all possible functions we expect to observe, is placed on $f$ as a Gaussian process $f(\mathbf{x}) \sim \mathcal{GP}(0,k(\mathbf{x},\mathbf{x}'))$ with zero mean and covariance function $k$ (usually defined by a radial-basis-function kernel): 
\begin{equation}
k(\mathbf{x},\mathbf{x}') = \tau^2 {\rm{exp}}(-\frac{1}{2} \sum_{j=1}^{j=d}\frac{(x_j-x_j')^2}{s_j^2} ), \label{RBF}
\end{equation}
where the amplitude $\tau^2$ and length scales $s_j$ are hyperparameters $\mathbf{\theta}=\{\tau, s_j\}$.

Following the Bayes' theorem, the posterior prediction for $f$ (corresponding to  \eqref{eq:post}) given the dataset $\mathcal{D}$ can be derived to be another Gaussian: 
\begin{equation}
    p(f(\mathbf{x})|\mathcal{D}) = \frac{p(f(\mathbf{x}),\mathbf{y})}{p(\mathbf{y})} = \mathcal{N}(\mu_{f}(\mathbf{x}|\mathcal{D}), {\rm{cov}}_{f}(\mathbf{x}, \mathbf{x}'|\mathcal{D})),
\label{eq:post_apd}
\end{equation}
with mean and covariance respectively:
 \begin{align}
    \mu_{f}(\mathbf{x}|\mathcal{D}) & = k(\mathbf{x}, \mathbb{X})^T {\rm{K}}(\mathbb{X},\mathbb{X})^{-1} \mathbf{y}, 
\label{eq:mean}\\
    {\rm{cov}}_{f}(\mathbf{x}, \mathbf{x}'|\mathcal{D}) & = k(\mathbf{x},\mathbf{x}') - k(\mathbf{x},\mathbb{X})^T {\rm{K}}(\mathbb{X},\mathbb{X})^{-1} k(\mathbf{x}',\mathbb{X}),
\label{eq:cov}
\end{align}
where matrix element ${\rm{K}}(\mathbb{X},\mathbb{X})_{ij}=k(\mathbf{x}^i,\mathbf{x}^j)$ and variance $\sigma^2_f(\mathbf{x}|\mathcal{D}) \equiv {\rm{cov}}_{f}(\mathbf{x}, \mathbf{x}|\mathcal{D})$. The hyperparameters $\mathbf{\theta}$ are determined which maximizes the likelihood function $p(\mathcal{D}|\mathbf{\theta})\equiv p(\mathbf{y}|\mathbf{\theta})=\mathcal{N}(0, {\rm{K}}(\mathbb{X},\mathbb{X}))$. 

For $f(\mathbf{x}) | \mathcal{D},\overline{y}(\tilde{\mathbf{x}})$ (to compute $h^{\pm}(l,a| \mathcal{D}, \overline{h}(\tilde{l},\tilde{a}))$ in \eqref{eq:u2}), we consider Bayes' theorem with $f(\mathbf{x}) | \mathcal{D}$ as a prior:
\begin{align}
    p(f(\mathbf{x})|\mathcal{D}, \overline{y}(\tilde{\mathbf{x}})) = 
    \frac{p(f(\mathbf{x}), \overline{y}(\tilde{\mathbf{x}})|\mathcal{D})}{p( \overline{y}(\tilde{\mathbf{x}})|\mathcal{D})}.
\end{align}
One can then get the mean and variance of $f(\mathbf{x})|\mathcal{D}, \overline{y}(\tilde{\mathbf{x}})$ using the standard conditional Gaussian formula: 
\begin{align}
    \mu_{f}(\mathbf{x}| \mathcal{D}, \overline{y}(\tilde{\mathbf{x}}))  = &  \mu_{f}(\mathbf{x}|\mathcal{D}) + \frac{\mathrm{cov}_{f}(\mathbf{x}, \tilde{\mathbf{x}}| \mathcal{D}) ( \overline{y}(\tilde{\mathbf{x}}) - \mu_{f}(\mathbf{x}|\mathcal{D}))}{\sigma^2_{f}(\mathbf{x}|\mathcal{D})}
\nonumber \\
     = &  \mu_{f}(\mathbf{x}|\mathcal{D}),
\\
    \sigma^2_{f}(\mathbf{x}| \mathcal{D}, \overline{y}(\tilde{\mathbf{x}})) = & \sigma^2_{f}(\mathbf{x}| \mathcal{D}) - \frac{\mathrm{cov}^2_{f}(\mathbf{x}, \tilde{\mathbf{x}}| \mathcal{D})}{\sigma^2_{f}(\tilde{\mathbf{x}}|\mathcal{D})}.
\label{eq:next}
\end{align}

\end{document}